\documentclass[%
reprint,
superscriptaddress,
 amsmath,amssymb,
 aps,
 pra,
]{revtex4-2}

\usepackage{nicefrac}
\usepackage{siunitx}
\usepackage{multirow}
\usepackage[normalem]{ulem}
\usepackage{graphicx}
\usepackage{dcolumn}
\usepackage{bm}

\newcommand{\figref}[1]{Fig.\,\ref{#1}}

\newcommand{\taburef}[1]{Tab.\,\ref{#1}}
\newcommand{\iso}[1]{$^{#1}$}
\newcommand{\fulltr}[2]{\ensuremath{3 d^{2} 4s \,^{4\!} F_{#1}\rightarrow 3 d^{2} 4p \,^{4\!} G_{#2}}}
\newcommand{\jVal}[2]{\ensuremath{\nicefrac{#1}{#2}}}
\newcommand{\tr}[2]{\ensuremath{\jVal{#1}{2}\rightarrow\jVal{#2}{2}}}

\DeclareSIUnit[]\Vpp
{\text{\ensuremath{\text{V}_{\textup{pp}}}}}

\def\orcid#1{\kern .08em\href{https://orcid.org/#1}{\includegraphics[keepaspectratio,width=0.7em]{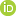}}}

\begin{document}


\title{Transition frequencies, isotope shifts, and hyperfine structure \\ in $4s \rightarrow 4p$ transitions of Ti$^+$ ions}

\author{Tim Ratajczyk}
\affiliation{Institut f\"ur Kernphysik, Technische Universit\"at Darmstadt, 64289 Darmstadt, Germany}%

\author{Kristian König\orcid{0000-0001-9415-3208}}
\email{kkoenig@ikp.tu-darmstadt.de}
\affiliation{Institut f\"ur Kernphysik, Technische Universit\"at Darmstadt, 64289 Darmstadt, Germany}%
\affiliation{Helmholtz Research Academy Hesse for FAIR, Campus Darmstadt, 64289 Darmstadt, Germany}

\author{Philipp Bollinger}
\affiliation{Institut f\"ur Kernphysik, Technische Universit\"at Darmstadt, 64289 Darmstadt, Germany}%

\author{Tim Lellinger} 
\affiliation{Institut f\"ur Kernphysik, Technische Universit\"at Darmstadt, 64289 Darmstadt, Germany}%

\author{Victor Varentsov} 
\affiliation{Facility for Antiproton and Ion Research in Europe (FAIR), Planckstr. 1, 64291 Darmstadt, Germany}%

\author{Wilfried N\"ortersh\"auser\orcid{0000-0001-7432-3687}}
\affiliation{Institut f\"ur Kernphysik, Technische Universit\"at Darmstadt, 64289 Darmstadt, Germany}%
\affiliation{Helmholtz Research Academy Hesse for FAIR, Campus Darmstadt, 64289 Darmstadt, Germany}

\author{Julien Spahn}
\affiliation{Institut f\"ur Kernphysik, Technische Universit\"at Darmstadt, 64289 Darmstadt, Germany}%



\date{\today}

\begin{abstract}
We have measured transition frequencies, isotope shifts and hyperfine structure splittings in the $3 d^{2}\left({ }^{3\!}F\right) 4 s\,{ }^{4} F_J\rightarrow 3 d^{2}\left({ }^{3\!} F\right) 4 p \,^{4} G_{J+1}$ transitions in Ti$^+$ ions for $J=\nicefrac{3}{2},\, \nicefrac{5}{2},\, \nicefrac{7}{2}$ using collinear laser spectroscopy. Ions were generated by laser ablation in a buffer gas atmosphere and extracted into vacuum through a nozzle and a pair of radiofrequency (RF) funnels. The obtained results are of interest as reference values for on-line measurements of short-lived titanium isotopes and for astrophysical searches for temporal or spatial variations of the fine structure constant $\alpha$ using quasar absorption spectra.       
\end{abstract}

\maketitle

\section{Introduction}
\label{sec:intro}
Measuring atomic transition frequencies and isotope shifts with high precision gives insights into atomic \cite{Sun.2020} and nuclear structure \cite{Otten.1989,Blaum.2013,Campbell.2016}, provides tests of QED \cite{Parthey.2011,Ullmann.2017,Micke.2020}, and is essential for atomic clocks \cite{Kozlov.2018}. Furthermore, those measurements can be used to determine masses of exotic particles, \textit{e.g.}, the pion \cite{Hori.2020}, to search for new interactions beyond the standard model \cite{Berengut.2018,Berengut.2020}, and for astrophysical studies that search for variations of the fine structure constant $\alpha$ \cite{Berengut.2011, Safronova.2014}.  
For the latter, titanium lines are of particular importance as those are a significant part of stellar spectra. While the spectrum of neutral titanium has been studied intensively in the past \cite{Channappa.1965,Aydin.1990,Stachowska.1993,Anastassov.1994,Luc.1996,Azaroual.1992,Jin.2009,Kobayashi.2019,Neely.2021}, only a small number of precision studies was performed in transitions of singly charged Ti$^+$ ions, which is the dominant ionization state in quasar absorbers.
Most of the investigated transitions are in the visible blue region \cite{BerrahMansour.1992,Nouri.2010} and only very few in the UV range \cite{Gianfrani.1991,Gangrsky.2004}. Theoretical analyses of the measured spectra and predictions of isotope shifts and hyperfine structure parameters have attracted increasing interest in recent years \cite{Bieron.1999,Berengut.2008,Bouazza.2013, Murphy.2013, Ruczkowski.2014}. 


The three transitions in Ti$^+$ ions studied in this work are of particular importance for astrophysical studies. From the isotope shifts, the isotope abundances in gas clouds in the early universe can be determined, models of the chemical evolution can be tested, and they are required for the search for variations of the fine structure constant $\alpha$ \cite{Berengut.2011}. 
Quasar absorption spectra allow one to look back in time in the universe. By comparing absorption lines in the quasar spectra with high precision measurements on earth, temporal changes can be extracted over billions of years and spatial variations in regions that are billions of light years apart \cite{Webb.1999,Berengut.2010}. Since the fine-structure splitting is proportional to $\alpha^2$, these measurements are a sensitive probe to temporal changes of the fine-structure constant \cite{Dzuba.1999}. Testing $\alpha$ for changes since the beginning of the universe allows one to determine whether we can rely on the natural constants being the same everywhere and anytime in the universe or if they change over time as some theories predict \cite{Uzan.2003}. The sensitivity on $\alpha$ varies widely and depends on the element, its charge state and transition. In a recent review, particularly sensitive transitions for the analysis of quasar absorption spectra were summarized \cite{Berengut.2011} and for some important transitions with a lack of experimental data, isotope shifts were calculated in a subsequent paper \cite{Murphy.2013} to support such analyses.
From these transitions, we have targeted the 338.47-nm transition in Ti$^+$ ions for which the isotope shift information is stated as ``extremely important'' and the absolute transition frequency as ``very important'' \cite{Berengut.2011} and measured both for all stable isotopes. 

A second motivation for our studies is the fact that Ti isotopes are in a key region to investigate nuclear structure due to the traditional shell closures at neutron numbers $N=20,28$ and the recently established shell closures at $N=32,34$ \cite{Wienholtz.2013,Steppenbeck.2013}. The elements in the proximity were a focus of recent collinear laser spectroscopy experiments \cite{Kreim.2014,Rossi.2015,Miller.2019,Koszorus.2021,GarciaRuiz.2016,Avgoulea.2011,Gangrsky.2004,Charlwood.2010,Heylen.2016,Minamisono.2016,Sommer.2022,Koenig.2023}.
Differential root mean square (rms) charge radii feature a kink at the $N=28$ shell closure for all isotopic chains measured so far \cite{Koszorus.2021,GarciaRuiz.2016,Charlwood.2010,Minamisono.2016,Sommer.2022}. A similar behavior is observed at all magic numbers that originate from the spin-orbit splitting ($N=28,\,50,\,82,\,126$) \cite{Ruiz.2020}. Contrary, at $N=20$ the rms charge radius is monotonically increasing without a significant kink in Ar, K and Ca \cite{Klein.1996,Behr.1997,Rossi.2015,Vermeeren.1996,Miller.2019}. In Sc, however, a kink at $N=20$ was recently observed but could not be described by theory \cite{Koenig.2023}. Measurements in neutron-deficient Ti can test if this kink manifests and is caused by different nuclear interactions by the additional proton in the $f_{\nicefrac{7}{2}}$ shell or, \textit{e.g.}, by continuum effects that occur at the proton dripline. 
In neutron-rich Ti isotopes, the neutron numbers $N=32$ and $N=34$ are of special interest, since those were found to be magic in Ca \cite{Wienholtz.2013,Steppenbeck.2013}. First charge radius measurements that reached and crossed the $N=32$ shell closure in Ca and K, respectively, however, do not exhibit signs of the characteristic kink \cite{GarciaRuiz.2016,Koszorus.2021} but further efforts to measure across $N=32$ and to $N=34$ are ongoing \cite{GarciaRuiz.2017}. 

As the discussed shell closures at $N=20,28,32,34$ are accessible in Ti, respective measurements will improve the understanding of nuclear structure in this region of the nuclear chart. Due to the refractive properties of Ti, the ion production of short-lived isotopes at ISOL facilities is challenging and so far only $^{44,45}$Ti were investigated \cite{Gangrsky.2004}.
In preparation for further collinear laser spectroscopy measurements in exotic Ti isotopes, optical transitions that are most sensitive with respect to the determination of differential rms charge radii and nuclear moments were identified in measurements of the stable isotopes in this work.
The Ti$^+$ ground state $3d^2(^3\mathrm{F})4s$ has nine possible fine-structure transitions to $3d^2(^3\mathrm{F})4p$ levels. The three $J\rightarrow J+1$ transitions starting at $J = \nicefrac{3}{2},\,\nicefrac{5}{2},\,\nicefrac{7}{2}$ are most promising as those feature the largest transition rates.
Here, we present isotope-shift measurements of $^{46-50}$Ti in all three transitions, providing accurate references for on-line measurements. Results are also combined with charge radii reported from the combination of elastic electron scattering and muonic atom spectroscopy \cite{Fricke.2004} to determine the field-shift and mass-shift parameters of the respective transitions. With those, nuclear charge radii can now be extracted from isotope shift measurements in exotic Ti isotopes.

\section{Experimental Setup}
\label{sec:setup}
The measurements were performed at the \textbf{Co}llinear \textbf{A}pparatus for \textbf{L}aser spectroscopy and \textbf{A}pplied Physics (COALA) at TU Darmstadt \cite{Konig.2020}. 
For this work, a new a helium-gas-buffered laser-ablation ion source was developed \cite{Ratajczyk.2020} that is described in detail in \cite{Ratajczyk.2022}. In brief, ion production takes place in a high-pressure (\SI{50}{mbar} helium) ablation stage and is realized with a short ($\SI{10}{ns}$) and intense (typically \SI{1}{\milli\joule}, $\sim \SI{100}{\micro\meter}$ diameter) pulse from a frequency-doubled, diode-pumped Q-switched Nd:YAG laser (Lumibird Centurion+) fired onto a titanium target surface with a repetition rate of \SI{100}{Hz}. The expansion of the hot plasma plume of released material is stopped by the surrounding helium gas. 
The generated ions are cooled by collisions with the buffer gas and transported with the gas stream through a supersonic nozzle, which connects the ablation stage with the adjacent funnel stage that operates at a gas pressure of slightly less than \SI{1}{mbar}. Here, the ions are guided through an RF-only funnel and the transversal diameter of the ion cloud is narrowed down to a diameter of \SI{1}{mm}. 
Finally, the ions, still dragged by the gas flow, pass a second nozzle into the third stage (buncher stage). Here, a DC gradient from an RF-DC ion guide funnel becomes responsible for the ion transport, since the gas pressure of \SI{1E-4}{mbar} is too low for ion transport. We note that the RF-DC ion guide can be used for the preparation of ion bunches. This was not applied during the Ti$^+$ measurements since the laser ablation itself produced temporal confined ion bunches with a length \SI{<1}{ms}, corresponding to the time it takes until all ions are extracted from the ablation region. As a last differential pumping barrier, a skimmer is installed at the transition to the ultra-high vacuum beam line. 
The complete ion source rests on a high-voltage platform that was held on 14\,kV with a stability of 0.01\,V by using a precision (5\,ppm absolute, $<1$\,ppm relative) voltage divider and a digital feedback loop \cite{Koenig.2024}. From here, the ions were electrostatically accelerated into the COALA beamline at ground potential.
The ion source delivered average ion currents of 10\,\si{\pA} measured on a Faraday cup, containing all stable isotopes of titanium according to their natural abundance.
With a $10^\circ$ bender, the ion beam was overlapped with a co- and a counterpropagating laser beam. 
Laser-ion interactions and the subsequent fluorescence detection was performed in a mirror-based optical detection region \cite{Maass.2020}.

The continuous-wave laser light used for spectroscopy was generated with a Ti-sapphire (Ti:Sa) laser (Matisse-2, Sirah Lasertechnik) at \SI{676}{\nm} pumped by a frequency-doubled Nd:YAG laser (Millennia eV, Spectra Physics). Since this wavelength is at the lower end of Ti:Sa operation, one of the Ti:Sa cavity mirrors was replaced by a mirror with low reflectivity for wavelengths $\lambda>700$\,\si{\nm} to efficiently suppress long-wavelength modes. The frequency of the Ti:Sa was stabilized on short timescales to a reference cavity, which was long-term stabilized to a wavelength meter (WS8, HighFinesse). 
The laser output was fed into the enhancement cavity of a frequency doubler (Wavetrain 2, Spectra Physics) equipped with an LBO crystal to generate \SI{338}{\nm} light. Spectroscopy was performed with laser powers of $150-300\,\mu$W.

When the Doppler-shifted laser frequency is in resonance with the ion transition, fluorescence photons are subsequently emitted and detected by photomultipliers. 
Instead of scanning the laser frequency, a voltage was applied to the fluorescence detection region (FDR) to change the ion velocity and, hence, the Doppler shift.
Spectra were recorded using the time-resolved data acquisition system TILDA \cite{Kaufmann.2015,Kaufmann.2019} based on two field-programmable gate arrays (FPGA), which produce a pattern for the device control, \textit{i.e.}, the scanning voltage, and record the time of each photomultiplier event in respect to the laser-ablation pulse. 

\begin{figure}[tbh]
        \centering
        \includegraphics[clip, trim=0mm 0mm 0mm 0mm,width=0.5\textwidth]{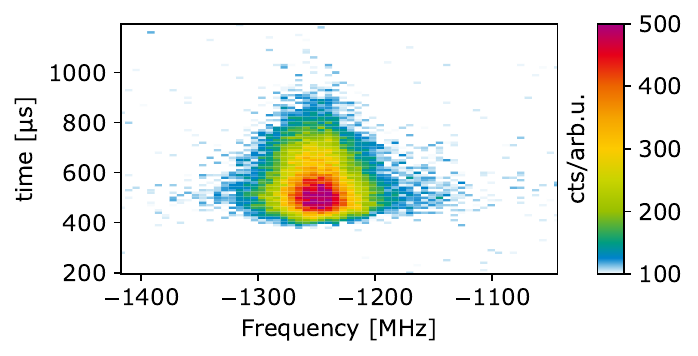}
        \includegraphics[clip, trim=0mm 0mm 0mm 0mm,width=0.5\textwidth]{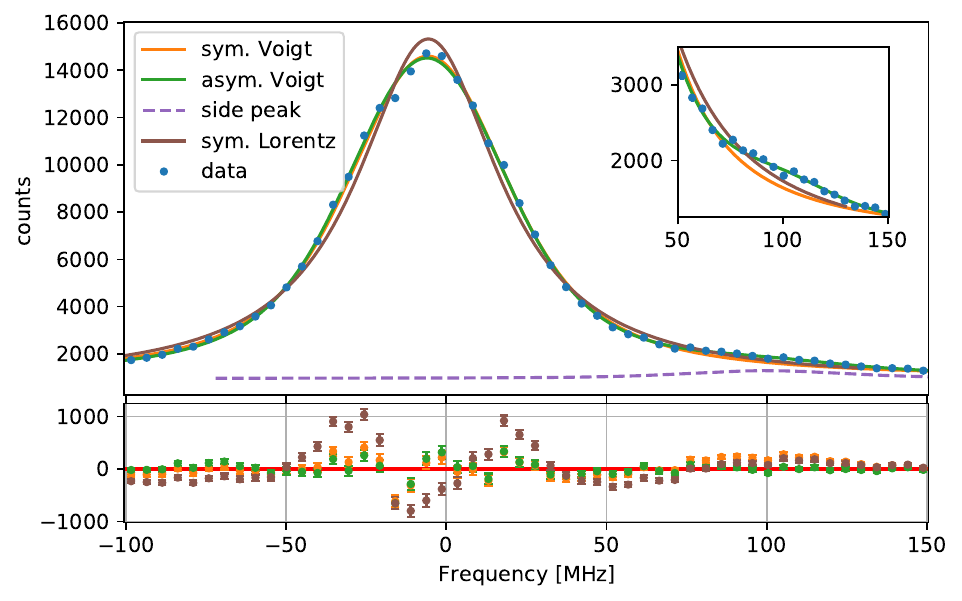}
        \caption{Top panel: Time-resolved spectrum of $^{48}$Ti in the \fulltr{\jVal{3}{2}}{\jVal{5}{2}}\ transition. The 2D-histogram shows the number of normalized photon events as a function of the laser frequency ($x$-axis, converted from Doppler tuning voltage) and the time elapsed since the ablation laser pulse ($y$-axis). Each bin corresponds to a voltage step of 255\,mV (6.32\,MHz) and a time of 10\,$\mu$s. The color coding is offset to fade out the relatively homogeneous background.\\
        Lower panel: Lineshape of $^{48}$Ti in the \tr{3}{5}\ transition fitted with several functions and fitting residuals in the lower trace. The Lorentzian profile (brown) shows the largest deviation while the Voigt and the asymmetric Voigt profile (= Voigt + side peak) describe the spectrum reasonably well. The latter performs better with respect to a small hump at higher energies that is enlarged in the inset. The additional side peak of the asymmetric Voigt at $\sim\SI{100}{MHz}$ is also depicted independently in the figure (purple).}
        \label{fig:lineshape}
\end{figure}

Such a time-resolved spectrum is shown in Fig.\,\ref{fig:lineshape}. The normalized count rate is color-coded. The signal appears to be symmetric along  the frequency ($x$) axis and extends over a time interval of about \SI{400}{\micro\s} on the $y$-axis with the main part of the intensity within the first \SI{200}{\micro\s}. The resonance lineshape is obtained by summing all events obtained at a specific frequency within a time window from 400--800\,\si{\micro\s}. 
Exemplary, a resonance curve in the \tr{3}{5}
transition is shown in the lower panel of Fig.\,\ref{fig:lineshape}. 

Several lineshape functions, \textit{i.e.}, a Lorentzian (brown), a Voigt (orange), and an asymmetric Voigt (green), were fitted to the resonance spectrum and the residua from the experimental data points are indicated in the lower trace. The asymmetric Voigt is constructed by adding a second smaller Voigt profile with the same linewidth parameters as the main peak at a $\approx 100$-MHz shifted frequency ($\approx 4.4$\,eV). Its contribution is also indicated (purple). 
Such side peaks in collinear laser spectroscopy of ions can arise from inelastic collisions with residual gas atoms that lead to electron excitations. The required energy is taken from the kinetic energy of the ion. In such a case, the ion energy is shifted by the excitation energy and those ions will appear shifted by this amount in the spectrum \cite{Krieger.2016}. The energy loss of the sidepeak observed here corresponds to a collisional excitation from the ground state into the 4-5 eV region, that offers an exceptionally high level density in Ti$^+$. 
In the analysis, all spectra were fitted with the three line profiles. With respect to the fitted center position, all fit functions agree within \SI{260}{kHz}, which is negligible compared to other systematical uncertainties that will be discussed below. Further analysis was therefore carried out using the symmetric Voigt profile. 

\section{Results and Discussion}
\label{sec:measurements}

    \begin{figure*}[tb]
        \centering
        \includegraphics[clip, trim=0mm 0mm 0mm 0mm,width=1\textwidth]{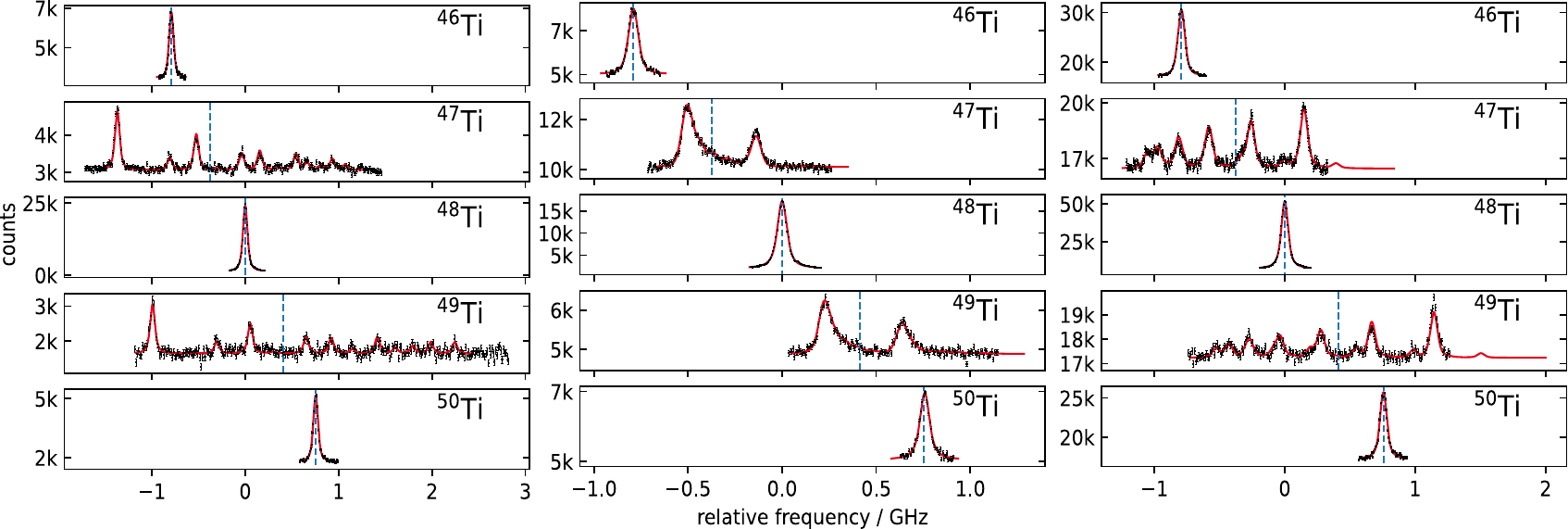}
        \caption{Spectra of the \fulltr{J}{J+1} transition for all five stable isotopes of Ti$^+$ (left: \tr{3}{5}, middle: \tr{5}{7}, right: \tr{7}{9}). Data points and corresponding statistical uncertainty are presented in black and a fit of the data with a symmetric Voigt profile in red. The blue vertical line marks the center of gravity. The frequency is shown relative to $^{48}$Ti. 
        }
        \label{fig:all_iso35}
    \end{figure*}
    

Example spectra of the three investigated \fulltr{J}{J+1}\ fine-structure transitions that were recorded in anticollinear geometry in the five stable isotopes are shown in \figref{fig:all_iso35}. 
The two odd isotopes have nuclear spins of $I=\nicefrac{5}{2}$~(\iso{47}Ti) and $I=\nicefrac{7}{2}$~(\iso{49}Ti), causing hyperfine splittings that are both fully resolved in the \tr{3}{5} transition, whereas they are largely collapsed into two broad lines in the \tr{5}{7} transition and partially resolved in the \tr{7}{9} transition. 
To extract the center-of-gravity frequencies and hyperfine parameters of the partially unresolved hyperfine structure, the peak intensity ratios were fixed to those expected from the Raccah coefficients and $^{47}$Ti and $^{49}$Ti were fitted with a common ratio of the upper and lower states' quadrupole hyperfine constants $B_u/B_l$. Since the hyperfine spectrum of the \tr{5}{7} transition is strongly collapsed, the $B_l(^{47}\mathrm{Ti})/B_l(^{49}\mathrm{Ti})$ ratio was additionally fixed to the ratio of the known electric quadrupole moments $Q(^{47}\mathrm{Ti})/Q(^{49}\mathrm{Ti})= 0.302/0.247$ \cite{Stone.2021}, which nicely agrees for the free fits in the \tr{3}{5} and \tr{7}{9} transitions.
The extracted hyperfine structure constants are listed in \taburef{tab:ABvals}. The \tr{3}{5} transition with the well separated hyperfine peaks yields, particularly for the $B$ parameters, a significantly higher precision for a similar measurement time and, hence, is favorable for the spectroscopy of short-lived isotopes in order to extract their nuclear moments.
For all transitions, the $A$ values are obtained with a precision at the 1-MHz level. In the absence of hyperfine structure anomaly, the ratio $A_i(^{47}\mathrm{Ti})/A_i(^{49}\mathrm{Ti})$ ($i\in \{ \mathrm{u,l}\}$) should be equal to the ratio of the respective nuclear $g_I$-factors of the two isotopes, which is $g_I(^{47}\mathrm{Ti})/g_I(^{49}\mathrm{Ti})=0.9997(2)$ \cite{Stone.2019}. Within the experimental uncertainty this is the case in all transitions, even in those with unresolved hyperfine structure. 
The even-parity configuration system of the lower states in Ti$^+$ has been theoretically analyzed and hyperfine $A$-factors were calculated for \iso{49}Ti in \cite{Bouazza.2013}, which are also included in \taburef{tab:ABvals}. Particularly, for the \tr{5}{7} and the \tr{7}{9} transition those agree well with the experimental results but deviate for the \tr{3}{5} transition by 20\,\%.

Another, semi-empirical approach calculated the hyperfine constants of Ti$^+$ for ground and excited states \cite{Ruczkowski.2014}. While the $A_u$ parameters are in reasonable agreement, the $A_l$ deviate significantly.
In the lowest fine-structure level ($J=\jVal{3}{2}$), the calculated $A$ is a factor of 2 too large while it is a factor $-12$ (wrong sign!) and 3 too small for the $J=\jVal{5}{2}$ and $\jVal{7}{2}$ levels, respectively, which can cause problems when using the calculated wavefunctions, \textit{e.g.}, for the determination of chemical abundances in stellar atmospheres. The calculated $B$ parameters are in reasonable agreement with the experimental results.

    \begin{table}[tb]
            \caption{Values of the hyperfine-structure $A$ and $B$ constants of the lower ($l$) and  upper ($u$) states in the three investigated \fulltr{J}{J+1} fine-structure transitions in units of MHz. Values in parentheses are the $1\sigma$ statistical uncertainties, which dominate over possible systematic contributions. Theoretical predictions for \iso{49}Ti by Ruczkowski \textit{et al}.\ \cite{Ruczkowski.2014} and Bouazza \cite{Bouazza.2013} are included.  }
        \label{tab:ABvals}
        \centering
        \begin{tabular}{llllll}
            \hline\hline
           &&$A_{l}$  &$A_{u}$  &$B_{l}$  &$B_{u}$  \\ \hline
            \multirow{4}{*}{$\tr{3}{5}$}
            & \iso{47}Ti & 68.2\,(6) & --120.6\,(5) & 19.5\,(2.4) & 74\,(5)    \\ 
            &\iso{49}Ti & 68.2\,(6) & --120.4\,(4)  & 16.2\,(2.1) & 50\,(6)  \\  
            &\iso{49}Ti\,\cite{Bouazza.2013} & 83.25 & ---  & ---  & --- \\
            &\iso{49}Ti\,\cite{Ruczkowski.2014}  & 174 & --126 & 15 & 58\\
                                 \hline
            \multirow{4}{*}{$\tr{5}{7}$}
            &\iso{47}Ti & --73.0\,(1.0) & --69.5\,(6)  & 23\,(15) & 89\,(17)  \\ 
            &\iso{49}Ti & --74.7\,(1.8) & --70.3\,(1.1)  & 19\,(12)  & 72(14) \\
            &\iso{49}Ti\,\cite{Bouazza.2013} & --72.57 & ---  & ---  & --- \\
            &\iso{49}Ti\,\cite{Ruczkowski.2014} & ~~6 & --79 & 15 & 56   \\  
                                \hline
            \multirow{4}{*}{$\tr{7}{9}$} 
            &\iso{47}Ti  & --114.3\,(1.2) & --43.5\,(9)  & 13\,(18) & 70\,(20) \\
            &\iso{49}Ti & --116.5\,(9)  & --45.1\,(7)  & 11\,(14) & 56\,(16)  \\
            &\iso{49}Ti\,\cite{Bouazza.2013} & --119.07 & ---  & ---  & --- \\
            &\iso{49}Ti\,\cite{Ruczkowski.2014} & --44 & --54  & 21 & 67 \\  \hline\hline
        \end{tabular}
    \end{table}
 


The measured isotope shifts $\delta\nu^{48,A} = \nu^{A} - \nu^{48}$ are listed in \taburef{tab:iso_shift}. The reported uncertainty is the geometrical average of the statistical uncertainty and a systematic uncertainty of \SI{\pm 600}{kHz}. The latter  
originates from the the uncertainty of the starting potential of the ions that might be offset to the electronically measured voltage due to contact voltages. 
The experimentally extracted isotope shifts are in excellent agreement with\textit{ ab initio} calculations that were performed in \cite{Murphy.2013} using the atomic structure package AMBIT that is an implementation of the combination of configuration interaction and many-body perturbation theory. The theoretical results are included in the last line of \taburef{tab:iso_shift} and differ by less than 6\,MHz.

\begin{figure}[tb]
    \centering
    \includegraphics[width=0.9\linewidth]{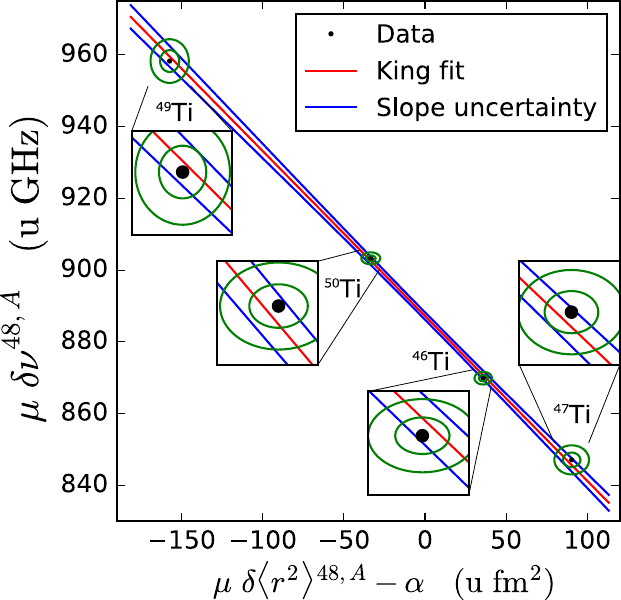}
    \caption{King fit with the determined modified isotope shifts $\mu\,\delta\nu^{48,A}$ with $\mu=m_{48}\cdot m_{A}/(m_{48}-m_{A})$ of the \tr{3}{5} transition and modified differential nuclear charge radii $\mu\,\delta\!\left\langle r^2 \right\rangle^{48,A}$ from \cite{Fricke.2004}. The green ellipses represent the 1$\sigma$ and 2$\sigma$ surrounding of the data points. The insets show an enlarged view of the position of the data point with respect to the fit line (red) and the fits 67\% confidence limit (blue). }
    \label{fig:KingFit}
\end{figure}

The measured isotope shifts allowed us to extract the field-shift factor $F$ and the mass-shift factor $K$ for the three transitions, which are crucial for determining charge radii from short-lived isotopes. The results are listed in \taburef{tab:FandK}. 
The $F$ parameter is determined by a King-Plot procedure as described in \cite{Gorges.2017}. A parameter $\alpha$ is used to shift the $x$-axis origin to a point were correlations between the slope $F$ and the intercept of the line (related to $K$) are eliminated. The corresponding King plots, exemplary shown for the \tr{3}{5} transition in Fig.\,\ref{fig:KingFit}, agree well with the expected linear slope, validating the obtained isotope shifts. Since the value for $K$ at the position $x=\alpha$ (indicated as $K_\alpha$) has no physical meaning, we additionally list the $K$ value extracted without $\alpha$ optimization in the table. Those can be compared with theoretical results that are available for the \tr{3}{5} transition in \cite{Murphy.2013}, yielding again an excellent agreement. While no theoretical uncertainty of $F$ was published, the experimental uncertainty of $K$ is significantly smaller than the result obtained by atomic theory.

Besides the measurements of the isotope shifts also the absolute transition frequencies were determined for the three transitions. For these measurements, an additional second laser was irradiated in collinear geometry, and spectroscopy was performed in fast alteration. This allowed us to deduce the rest-frame frequency $\nu_0^2=\nu_\mathrm{col} \cdot \nu_\mathrm{acol}$ from the resonant collinear ($\nu_\mathrm{col}$) and anticollinear laser frequency ($\nu_\mathrm{acol}$) and removed the systematic uncertainty of the starting potential. The dominant remaining systematic contributions were the frequency measurement with the applied WS8 wavelength meter of 3.3\,MHz (1$\sigma$) and the spatial overlap of ion and laser beams, which was estimated to be 2\,MHz as in \cite{Imgram.2023.PRA}. The results are listed in Tab.\,\ref{tab:absolut_freq} together with literature values. The uncertainties of the rest-frame transition frequencies were improved by factors 6 to 15 compared to the previous most accurate measurements and are now sufficiently accurate to evaluate astrophysical observations \cite{Berengut.2011}.
Our measurements confirm the results published by Morton \cite{Morton.2003} and Saloman \cite{Saloman.2012} as well as the measurements of Aldenius \cite{Aldenius.2006}. Particularly the values from \cite{Morton.2003,Aldenius.2006} deviate by less than 1$\sigma$, while the reevaluation of the latter by Nave \cite{Nave.2012} actually shifted the transition frequency further away and displays a 1.6$\sigma$ discrepancy to our value. This is an important observation since Refs.\,\cite{Nave.2011,Nave.2012} recalibrated the frequency scales of several previous Fourier-Transform Spectroscopy measurements to establish a consistent frequency scale for a large subset of metal-ion transitions that are now used for varying-$\alpha$ analyses. According to our results, care should be taken when using these recalibrated values. Further measurements in other transitions should be carried out to consolidate this issue. 

\begin{table}[htb]
        \caption{Isotope shift $\delta\nu^{48,A} = \nu^{A} - \nu^{48}$ of all stable titanium isotopes with respect to $^{48}$Ti in the three investigated $3 d^{2} 4s \,^{4\!} F_J\rightarrow 3 d^{2} 4p \,^{4\!} G_{J+1}$ fine-structure
        transitions. The uncertainties are obtained from the statistical uncertainty of the fits and a systematic uncertainty of \SI{600}{\kHz} added in quadrature. 
        }
            \centering
            \begin{tabular}{lcccc} \hline
                 Transition & \iso{46}Ti & \iso{47}Ti&\iso{49}Ti&\iso{50}Ti  \\ \hline 
                  $\nicefrac{3}{2} \rightarrow \nicefrac{5}{2}$ & -788.1\,(0.8) & -375.0\,(0.9) & 408.5\,(1.3) & 753.6\,(0.7)\\ 
                 $\nicefrac{5}{2} \rightarrow \nicefrac{7}{2}$ & -790.2\,(1.0) & -373.6\,(2.0) & 412.4\,(2.2) & 754.4\,(0.7)\\ 
                 $\nicefrac{7}{2} \rightarrow \nicefrac{9}{2}$ & -792.6\,(0.7) & -377.9\,(1.3) & 408.7\,(1.6) & 756.9\,(1.0) \\ \hline

                 $\nicefrac{3}{2} \rightarrow \nicefrac{5}{2}$ \cite{Murphy.2013} & -794.15 & -377.74 & 405.32 & 753.68\\  \hline 
            \end{tabular}
            \label{tab:iso_shift}
    \end{table}
   
    \begin{table}[tbh]
        \caption{Extracted field shift and mass shift constants from the King Plot analysis of the three measured transitions. }
        \centering
        \begin{tabular}{lcccc} \hline
              Transition&$F$ [MHz/fm$^2$]& $K$ [GHz/u]&$K_\alpha$ [GHz/u]&$\alpha$ [ u fm$^2$] \\ \hline\hline
              $\nicefrac{3}{2} \rightarrow \nicefrac{5}{2}$&-460\,(34)&818.3\,(5.4)&886.8\,(2.1)&-149\\
              $\nicefrac{5}{2} \rightarrow \nicefrac{7}{2}$&-490\,(40)&881.5\,(6.4)&889.0\,(2.3)&-150\\
              $\nicefrac{7}{2} \rightarrow \nicefrac{9}{2}$&-439\,(34)&825.9\,(5.4)&890.8\,(2.1)&-148 \\ \hline

              $\nicefrac{3}{2} \rightarrow \nicefrac{5}{2}$ \cite{Murphy.2013} & -408 &  832\,(104) & &\\ \hline
              
        \end{tabular}
        \label{tab:FandK}
    \end{table}
    
    \begin{table*}[tbh]
        \caption{Transition frequencies compared with literature values. The transitions were first measured by Huldt \cite{Huldt.1982} and later improved by Zapadlick, whose results are only published as a private communication in Morton \cite{Morton.2003} and Saloman \cite{Saloman.2012}. Aldenius \cite{Aldenius.2006}  remeasured  the $J = \nicefrac{3}{2} \rightarrow \nicefrac{5}{2}$ transition which was later reevaluated by Nave \cite{Nave.2012}.}
            \centering
            \begin{tabular}{llll} \hline
                  & $J = \nicefrac{3}{2} \rightarrow \nicefrac{5}{2}$& $J = \nicefrac{5}{2} \rightarrow \nicefrac{7}{2}$ &$J = \nicefrac{7}{2} \rightarrow \nicefrac{9}{2}$ \\ \hline\hline
                  Huldt \cite{Huldt.1982}&885\,717\,630\,(5000)&888\,598\,636\,(5000)&891\,657\,418\,(5000)\\
                  Morton \cite{Morton.2003}&885\,720\,359\,(90)&888\,600\,075\,(90)&891\,661\,526\,(90)\\ 
                 Saloman \cite{Saloman.2012}&-&888\,600\,045\,(90)&891\,661\,496\,(90)\\ 
                 Aldenius \cite{Aldenius.2006}&885\,720\,448\,(30)& - &-\\ 
                 Nave \cite{Nave.2012}&885\,720\,478\,(30)&-&-\\
                 This work&885\,720\,428\,(5)& 888\,600\,144\,(6) &891\,661\,616\,(6)\\ \hline
            \end{tabular}
            \label{tab:absolut_freq}
    \end{table*}

\section{Conclusion}
\label{sec:conclusion}
We reported on first measurements using a new laser-ablation ion-source at COALA that was employed to create a beam of stable Ti ions for collinear laser spectroscopy. Three transitions from the ground state quadruplet were measured for all five stable isotopes. The rest-frame frequencies were determined with an uncertainty of better than \SI{6}{\MHz}, corresponding to a relative accuracy of about $7\times 10^{-9}$, and the isotope shifts were extracted with 1-MHz accuracy. The obtained results can be used to reference astrophysical quasar measurements. The new versatile ion source will allow us to perform further reference measurements, which are important for astrophysics  \cite{Berengut.2011, Murphy.2013} as well as for atomic and nuclear physics. 

For the stable odd-mass isotopes $^{47,49}$Ti, the hyperfine $A$ and $B$ factors of all involved atomic levels were extracted from the spectra. The \tr{3}{5} transition offered the largest splitting and, hence, the best resolution, which makes it a good candidate for the investigation of charge radii and nuclear moments of short-lived nuclei. The precise $A$ and $B$ parameters can serve as a reference to extract nuclear electromagnetic moments. Moreover, the field-shift factor $F$ and mass-shift factor $K$, extracted in this work, will allow one to determine nuclear charge radii from isotope-shift measurements in exotic Ti$^+$ isotopes to investigate the nuclear structure around the $N=20$ and the $N=32,34$ shell closures in neutron-deficient and neutron-rich titanium.

\begin{acknowledgments}
We acknowledge financial support from the German Federal Ministry for Education and Research (BMBF) under Contract nos.\ 05P19RDFN1 and 05P21RDFN1, the Deutsche Forschungsgemeinschaft (DFG, German Research Foundation) -- Project-ID 279384907 -- SFB 1245, and under Grant INST No. 163/392-1 FUGG. T.R.\ acknowledges support from HGS-HIRe.
\end{acknowledgments}

\bibliography{Literature.bib}

\end{document}